\documentclass[12pt]{iopart}

\usepackage{graphicx}
\usepackage{iopams}
\usepackage{color}
\usepackage{cite}

\newcommand{\thetaf}[4]{%
\theta\!\left[ {#1}\atop{#2} \right]\!\left(#3,#4\right)}

\begin{document}

\title[Quasielectrons as inverse quasiholes in lattice FQH models]{Quasielectrons as inverse quasiholes in lattice fractional quantum Hall models}

\author{Anne E. B. Nielsen$^{1,2,3}$, Ivan Glasser$^2$, Iv\'an D. Rodr\'iguez$^2$}
\address{$^1$ Max-Planck-Institut f{\"u}r Physik komplexer Systeme, D-01187 Dresden, Germany}
\address{$^2$ Max-Planck-Institut f{\"u}r Quantenoptik, D-85748 Garching, Germany}
\address{$^3$ Department of Physics and Astronomy, Aarhus University, DK-8000 Aarhus C, Denmark}

\begin{abstract}
From an experimental point of view, quasielectrons and quasiholes play very similar roles in the fractional quantum Hall effect. Nevertheless, the theoretical description of quasielectrons is known to be much harder than the one of quasiholes. The problem is that one obtains a singularity in the wavefunction if one tries to naively construct the quasielectron as the inverse of the quasihole. Here, we demonstrate that the same problem does not arise in lattice fractional quantum Hall models. This result allows us to make detailed investigations of the properties of quasielectrons, including their braiding statistics and density distribution on lattices on the plane and on the torus. We show that some of the states considered have high overlap with certain fractional Chern insulator states. We also derive few-body Hamiltonians, for which various states containing quasielectrons are exact ground states.
\end{abstract}

\section{Introduction}

While elementary particles are either bosons or fermions, it has turned out to be possible in certain complex quantum many-body systems to create anyonic quasiparticles with different and very peculiar properties. Anyons can, e.g., have a charge that is only a fraction of the elementary charge, and if two anyons are exchanged, it leads to a change of the wavefunction that is not just multiplication by a plus or a minus sign. Given that it has a huge impact on the properties of a system, whether the constituent particles are bosons or fermions, it is no surprise that anyons are attracting much attention.

Anyons can appear in connection with the fractional quantum Hall effect, which has been realized in certain two-dimensional structures at high magnetic field and low temperature. Various fractional quantum Hall states are obtained for different values of the magnetic field, and anyonic quasiholes/quasielectrons with positive/negative charge can be added to these states by slightly increasing/decreasing the magnetic field. From an experimental point of view, quasiholes and quasielectrons hence play very similar roles, but nevertheless it has turned out to be much harder to describe quasielectrons theoretically. As explained, e.g., in \cite{MRqe3}, this is because it is easier to modify the wavefunction to reduce the electron density locally than to increase it due to the Pauli exclusion principle.

One can introduce quasiholes in the wavefunction by inserting flux tubes with positive flux, but if one instead inserts a flux tube with negative flux to get a quasielectron, one gets a singularity in the wavefunction \cite{laughlin,laughlin-book}. This has triggered a lot of work to find suitable wavefunctions of quasielectrons and also the antiparticles of these quasielectrons \cite{laughlin,laughlin-book,MR,jainPRB2003,PhysRevB.76.075347,MRqe1, MRqe2,MRqe3,suorsa,MRqe4,MRqe5,PhysRevLett.112.026804,torus}. Laughlin proposed a quasielectron wavefunction already in his original paper on the Laughlin state \cite{laughlin}. A different wavefunction with a lower variational energy based on the composite fermion approach was later introduced by Jeon and Jain \cite{jainPRB2003}. The construction of quasielectron wavefunctions and of their antiparticles has shown a way to construct all the states in the Abelian quantum Hall hierarchy by consecutive condensations of quasielectrons and quasiholes \cite{suorsa}. Quasielectron wavefunctions have also been constructed for Moore-Read states \cite{MRqe1,MRqe2,MRqe3,MRqe4,MRqe5}, and recently it has been shown how to obtain a state for a Laughlin quasielectron on the torus \cite{torus}. In general, however, the obtained quasielectron wavefunctions are significantly more complicated than the simple quasihole wavefunctions, which makes it difficult to compute properties of quasielectrons.

In the last years, there has been significant progress for explicitly computing braiding properties of various types of quasiholes \cite{arovas,paredes,bonderson,kapit,wu,nielsen,liu}, but the quasielectrons are again more complicated to work with. For the latter it has been found that the braiding properties are as expected for the composite fermion version of the Laughlin quasielectrons \cite{jeon1,jeon2}, while for Laughlin's proposal for the quasielectron wavefunction the expected braiding statistics is obtained in the thermodynamic limit when considering a finite size braiding loop \cite{kjonsberg,JeonPRB2010}.

Here, we propose to solve the problem of complicated quasielectron wavefunctions and lack of symmetry between quasielectrons and quasiholes by putting the fractional quantum Hall system on a lattice. Lattice versions of fractional quantum Hall models are interesting for more reasons. One of the big aims at the moment within the field of ultracold atoms in optical lattices is to realize lattice fractional quantum Hall models and observe anyons \cite{Gross995}. The lattice gives rise to interesting features that are not present in the continuum systems \cite{liureview}, and the models may pave the way to realize fractional quantum Hall physics in new settings, possibly even at room temperature \cite{wen}.

In a lattice, the electrons can only be placed at a given set of lattice sites rather than in a continuous space, and this means that the singularity does not appear. It is even possible to define lattice models in such a way that there is an exact symmetry between quasielectrons and quasiholes. A consequence of this symmetry is that the charge distribution of a quasielectron is minus the charge distribution of a quasihole. In this construction, the properties of the quasielectrons are not more complicated to compute than those of the quasiholes. In addition, the wavefunctions are in many cases suitable for Monte Carlo simulations, so that systems with hundreds of lattice sites can be studied. We use this below to compute the shape and braiding properties of quasielectrons.

Another important question is whether the states constructed are related to ground states of fractional quantum Hall Hamiltonians on the lattice, such as fractional Chern insulators \cite{neupert,bernevig,liureview}. To investigate this, we consider a particular fractional Chern insulator model and find that the ground states of this model have a high overlap with the analytical wavefunctions. We also derive few-body Hamiltonians for which particular states containing quasielectrons are ground states. These Hamiltonians provide a convenient framework for manipulating anyons, since they allow both creation, braiding, and fusion of the anyons by changing the interaction strengths in the Hamiltonian.

\section{Wavefunctions}

Let us first investigate the case of lattice Laughlin states with $1/q$, $q\in\mathbb{N}$, particles per flux unit in a disc geometry. Specifically, we consider a 2D lattice with $N$ sites at the positions $(\textrm{Re}(z_j),\textrm{Im}(z_j))$ and take the local Hilbert space on site $j$ to be $|n_j\rangle$, where the occupation number $n_j$ can be either $0$ (empty site) or $1$ (occupied site). We also allow $K$ quasiholes with charges $p_k/q$, $p_k\in\mathbb{N}$, to be present at the positions $(\textrm{Re}(w_k),\textrm{Im}(w_k))$. The family of lattice Laughlin states is then defined as
\begin{equation}\label{state}
\fl |\psi\rangle_{\vec{p}}
=\mathcal{C}^{-1}\sum_{n_1,\ldots,n_N}\delta_{n}\prod_i\chi_{n_i}
\prod_{i,j}(w_i-z_j)^{p_in_j}
\prod_{i<j}(z_{i}-z_{j})^{qn_{i}n_{j}-n_i\eta-n_j\eta}
|n_1,\ldots,n_N\rangle,
\end{equation}
where $\vec{p}\equiv(p_1,p_2,\ldots,p_K)$, $\mathcal{C}$ is a real normalization constant, $\delta_{n}$ is unity if the number of particles is $\sum_jn_j=(N\eta-\sum_ip_i)/q$ and zero otherwise, $\chi_{n_j}= e^{i\phi_0+i\phi_jn_j}$ (with $\phi_0,\phi_j\in\mathbb{R}$) are unspecified single particle phase factors, and $\eta$ is a parameter that determines the number of lattice sites per area. In the following, we scale the lattice such that the area per lattice site is $2\pi\eta$, which corresponds to setting the magnetic length to unity.

The reason why the states \eref{state} are suitably called lattice Laughlin states is that they, for a suitable choice of the phase factors $\chi_{n_i}$, only differ from the normal Laughlin states in the continuum with $q$ odd by restricting both the allowed positions of the particles and the so-called neutralizing background charge to be on the chosen lattice. This can be seen by noting \cite{contlim} that  $|\psi\rangle\propto\sum_{n_1,\ldots,n_N}\langle 0|W_{p_1}\cdots W_{p_K} V_{n_1}V_{n_2} \cdots V_{n_N}|0\rangle|n_1,\ldots,n_N\rangle$, where
\begin{equation}
V_{n_j}=\chi_{n_j}:e^{i(qn_j-\eta)\phi(z_j)/\sqrt{q}}:,\qquad\label{vopV}
W_{p_k}={}:e^{ip_k\phi(w_k)/\sqrt{q}}:,\label{vopW}
\end{equation}
and comparing to the corresponding results for the continuum Laughlin states in \cite{MR}. Here $|0\rangle$ is the vacuum state, $:\ldots:$ means normal ordering, and $\phi(z_j)$ is the chiral field of a free, massless boson. In addition, numerical studies \cite{contlim,lp} for $q\leq6$ show that the topological properties of the states typically remain after introducing the lattice (the square lattice at $\eta=q/2$ and $q\geq5$ is an exception).

Our claim in the present work is that we can obtain wavefunctions for states with quasielectrons simply by taking some or all of the $p_i$ in \eref{state} to be negative integers. If, in addition, we take $\eta=q/2$ (corresponding to that the lattice filling factor $\sum_i n_i/N$ is $1/2$ in the absence of anyons), there is a symmetry, which ensures that the quasielectrons have exactly the same shape as the quasiholes. We will now justify these statements by computing the charge, density profile, and braiding statistics of the quasielectrons.

\section{Quasielectron charge and density profile}

The condition $\sum_jn_j=(N\eta-\sum_ip_i)/q$, coming from the $\delta_n$ factor in the wavefunction, shows that if we add a quasielectron to the system, then the total number of particles in the system increases by $-p_i/q$ independent of $\eta$. In the fractional quantum Hall effect, the particles are electrons with charge $-1$, and the quasielectrons hence have the expected charge $p_j/q$.

\begin{figure}
\begin{indented}\item[]
\includegraphics[width=10cm,bb=113 208 557 618,clip]{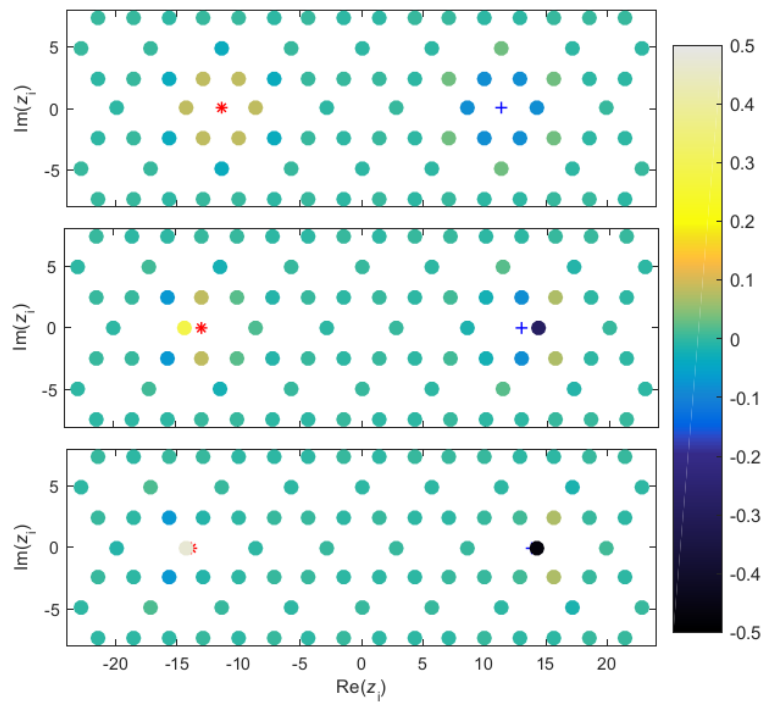}
\caption{Modification of the particle density due to the presence of anyons in the lattice Laughlin state \eref{state} at $q=3$ and half lattice filling ($\eta=q/2$). The lattice is chosen to be a kagome lattice defined on a disc with radius $27.9$. A quasielectron (quasihole) with charge $-1/3$ ($+1/3$) is placed at the position \textcolor{red}{$*$} (\textcolor{blue}{$+$}), and the color of the $j$th lattice site shows $\langle n_j\rangle_{(-1,+1)}-\langle n_j\rangle_{(0,0)}$.\label{fig:plane}}
\end{indented}
\end{figure}

Let us now take $\eta=q/2$ (i.e., lattice filling $\sum_jn_j/N=1/2-\sum_ip_i/(Nq)$). In this case, the coefficients of the state \eref{state} are invariant under the transformation
\begin{equation}
n_i\to 1-n_i  \quad \& \quad
p_i\to -p_i
\end{equation}
up to single particle phase factors (or up to a global phase factor if we choose $\chi_{n_j}$ such that $\chi_{n_j}\propto\chi_{1-n_j}$). Note that this transformation transforms quasiholes into quasielectrons and \textit{vice versa}. In particular, it follows that $\langle n_j\rangle_{\vec{p}}=1-\langle n_j\rangle_{-\vec{p}}$ and also $\langle n_j\rangle_{\vec{0}}=1-\langle n_j\rangle_{\vec{0}}$, so that
\begin{equation}\label{density}
\langle n_j\rangle_{\vec{p}}-\langle n_j\rangle_{\vec{0}}
=-(\langle n_j\rangle_{-\vec{p}}-\langle n_j\rangle_{\vec{0}}) \quad \textrm{for }\eta=q/2.
\end{equation}
It is natural to define the density profile of an anyon to be the difference between $\langle n_j\rangle$ when the anyon is present and when the anyon is not present. Equation \eref{density} then says that a quasihole with charge $p_i/q$ has the same density profile as a quasielectron with charge $-p_i/q$ except for a sign. This result is illustrated in Fig.~\ref{fig:plane} for different positions of the anyons on a kagome lattice. The figure also shows how the shape of the density profile varies with position. We note that even if a quasihole (quasielectron) approaches a lattice site, no singularity occurs. All that happens in that limit is that the probability that the site is empty (occupied by one particle) approaches unity.

\begin{figure}
\begin{indented}\item[]
\includegraphics[width=10cm,bb=69 208 555 596,clip]{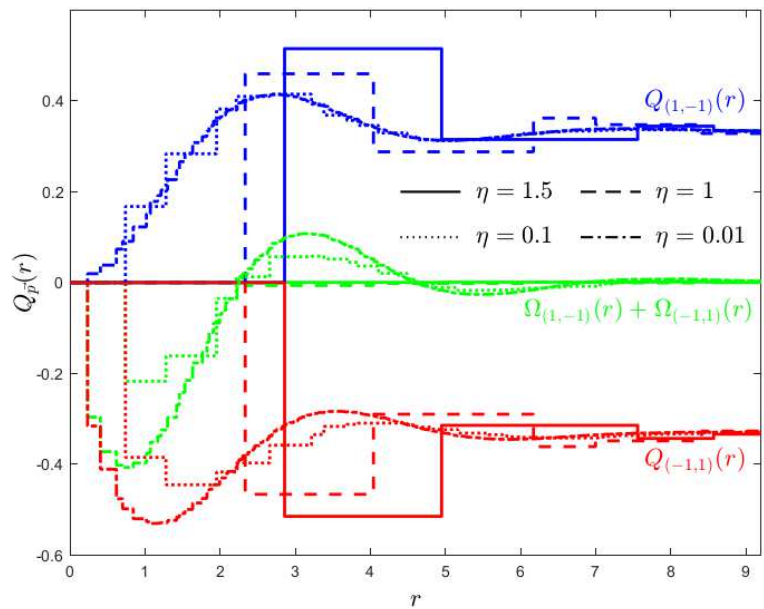}
\caption{Excess charge \eref{excharge} of a quasihole/quasielectron (blue/red) and their sum (green) in the lattice Laughlin state \eref{state} with $q=3$ for different densities of the lattice sites. The quasihole/quasielectron is placed at the origin at the center of a hexagon in a kagome lattice. The lattice is defined on a disc with radius $27.9$ for $\eta=1.5,1,0.1$ and $18.2$ for $\eta=0.01$, and an anyon of the opposite charge is placed at infinity.\label{fig:excharge}}
\end{indented}
\end{figure}

What happens if we move away from the symmetric point $\eta=q/2$? Figure~\ref{fig:excharge} shows the excess charge
\begin{equation}\label{excharge}
Q_{\vec{p}}(r)=-\sum_{\{i\in\{1,2,\ldots,N\} {\textstyle|} |z_i-w_1|\leq r\}}(\langle n_i\rangle_{\vec{p}}-\langle n_i\rangle_{\vec{0}})
\end{equation}
for $q=3$ and different values of $\eta$, and we observe that the curves for quasiholes and for quasielectrons are close to being symmetric for a broad range of $\eta$ values. If, however, we let $\eta$ approach zero (corresponding to that we approach the continuum limit), we observe that the excess charge of the quasihole converges to a fixed curve, while the charge distribution of the quasielectron becomes more and more narrow. This is because the singularity in the continuum wavefunction is approached. Figure~\ref{fig:excharge} also shows that the radius of the quasihole is only a bit larger in the lattice than it is in the continuum, which is in line with the results in \cite{liu}.

\section{Braiding statistics}

The Berry phase $\theta_k=i\oint_c\langle\psi|\frac{\partial \psi}{\partial w_k}\rangle dw_k+c.c.$ \cite{berry1,berry2} of \eref{state} obtained when moving the $k$th anyon around a closed curve $c$ evaluates to
\begin{equation}\label{braid}
\theta_k=i\frac{p_k}{2}\oint_c\sum_i\frac{\langle n_i\rangle_{\vec{p}}}{w_k-z_i}dw_k+c.c.,
\end{equation}
where we have assumed that $\chi_{n_i}$ is chosen such that it does not depend on $w_k$. To find the statistics of the anyons we need to compute $\theta_{k,(w_j\textrm{ inside})}-\theta_{k,(w_j\textrm{ outside})}$, where $\theta_{k,(w_j\textrm{ inside})}$ ($\theta_{k,(w_j\textrm{ outside})}$) is the Berry phase when the $j$th anyon is inside (outside) $c$ and not close to $c$. Our numerical results for $q=3$ in Figs.~\ref{fig:plane} and \ref{fig:excharge} (and for $q=2$ and $q=4$, not shown) show that the anyons only alter the particle density in a small region close to $w_j$. As long as we keep the anyons well separated, we hence have that $\langle n_i\rangle_{\vec{p},(w_j\textrm{ inside})}-\langle n_i\rangle_{\vec{p},(w_j\textrm{ outside})}$ is only nonzero close to the two possible positions of the $j$th anyon and furthermore does not depend on $w_k$. This allows us to move it outside the integral. Utilizing $\sum_{(i\textrm{ inside }c)}(\langle n_i\rangle_{\vec{p},(w_j\textrm{ inside})}-\langle n_i\rangle_{\vec{p},(w_j\textrm{ outside})})=-p_j/q$ for $w_j$ not close to $c$, we arrive at $\theta_{k,(w_j\textrm{ inside})}-\theta_{k,(w_j\textrm{ outside})}=2\pi p_j p_k/q$, which is the expected statistics for Laughlin anyons with charges $p_j/q$ and $p_k/q$.

\section{Quasielectrons on the torus}

\begin{figure}
\begin{indented}\item[]
\includegraphics[width=10cm,bb=73 227 552 592,clip]{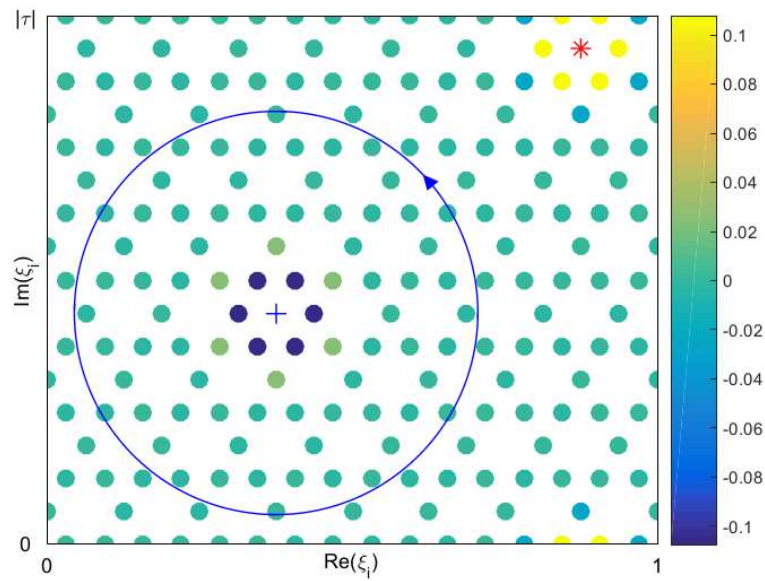}
\caption{Quasihole (\textcolor{blue}{$+$}) and quasielectron (\textcolor{red}{$*$}) on the torus in the state $|\psi_0\rangle_{(1,-1)}$ (see \eref{statetorus}) with $q=2$. The color of the lattice sites shows $\langle n_j\rangle_{(1,-1)}-\langle n_j\rangle_{(0,0)}$. When moving the quasihole around the blue curve we find numerically, using Monte Carlo simulations, that the difference in Berry phase when the quasielectron is at \textcolor{blue}{$+$} and at \textcolor{red}{$*$}, respectively, is $\phi=-3.145$ with a statistical error of order $0.003$. This is in agreement with the expected result $-\pi$.\label{fig:torus}}
\end{indented}
\end{figure}

The mathematics of fractional quantum Hall states defined on a torus is more complicated \cite{PhysRevB.31.2529,PhysRevLett.55.2095}, and it is therefore particularly challenging to construct quasielectrons on the torus. In fact, this problem has been solved only recently for Laughlin quasielectrons in continuous systems \cite{torus}. Utilizing the above ideas, however, the construction of quasielectrons in lattice systems on the torus is not more complicated than the construction of quasiholes. We define a torus by specifying the complex numbers $r_1$ and $r_2$ and identifying all points in the complex plane separated by $nr_1+mr_2$ with $n,m\in\mathbb{Z}$. Without loss of generality we shall take $r_1$ real and assume that $\textrm{Im}(r_2)>0$. We define $\tau=r_2/r_1$, $\xi_j=z_j/r_1$, and $\zeta_k=w_k/r_1$. It is then straightforward to generalize the derivation of lattice Laughlin states on the torus in \cite{deshpande} to include quasiholes and quasielectrons:
\begin{eqnarray}\label{statetorus}
\fl|\psi_{l}\rangle_{\vec{p}}=\mathcal{C}^{-1}\sum_{n_1,\ldots,n_N}\delta_{n}\prod_i\chi_{n_i} \prod_{i,j}\thetaf{1/2}{1/2}{\zeta_i-\xi_j}{\tau}^{p_in_j}\nonumber\\
\times \thetaf{a+l/q}{b}{\sum_{i=1}^N \xi_i (qn_i - \eta)+\sum_{j=1}^K\zeta_jp_j}{q\tau}\nonumber\\
\times\prod_{i<j}\thetaf{1/2}{1/2}{\xi_i-\xi_j} {\tau}^{qn_in_j-n_i\eta-n_j\eta}|n_1,\ldots,n_N\rangle.
\end{eqnarray}
Here, $\thetaf{a}{b}{\xi}{\tau}\equiv\sum_{n \in \mathbb{Z} } e^{i\pi \tau(n+a)^2 + 2\pi i (n+a)(\xi + b)}$ and $l\in\{0,1,\ldots,q-1\}$. For $q$ even $(a,b)=(0,0)$, and for $q$ odd $(a,b)$ can be either $(0,0)$, $(0,1/2)$, $(1/2,0)$, or $(1/2,1/2)$. For $\eta=q/2$ and $l=0$ or $l=q/2$, the derivation leading to Eq.~\eref{density} is again valid, and the density profiles for quasielectrons and quasiholes are minus each other. For other values of $l$, we need to add $l\to q-l$ to the transformation $n_i\to1-n_i$ and $p_i\to-p_i$. Since states with different $l$ are connected through spectral flow, and since spectral flow should not alter the density profile of localized anyons, this should, however, not change the picture. In Fig.~\ref{fig:torus} we show numerically for $q=2$ that the anyons are localized and obey the expected statistics.

\section{Comparison of the anyon wavefunctions with fractional Chern insulator wavefunctions}

Having established the topological properties of the states, a natural next question is whether these states are related to ground states of fractional quantum Hall Hamiltonians on the lattice. We shall here consider the model studied in \cite{Sorensen,Hafezi}, where it was shown that a bosonic fractional quantum Hall state with $q=2$ can be realized on a square lattice using a Bose-Hubbard model with Peierls substitution and interactions, provided the lattice filling factor $\eta/2$ is less than about $0.2$. This model is particularly relevant, since it allows the lattice filling factor to be adjusted and has the correct continuum limit.

\begin{figure}
\begin{indented}\item[]
\includegraphics[scale=0.60]{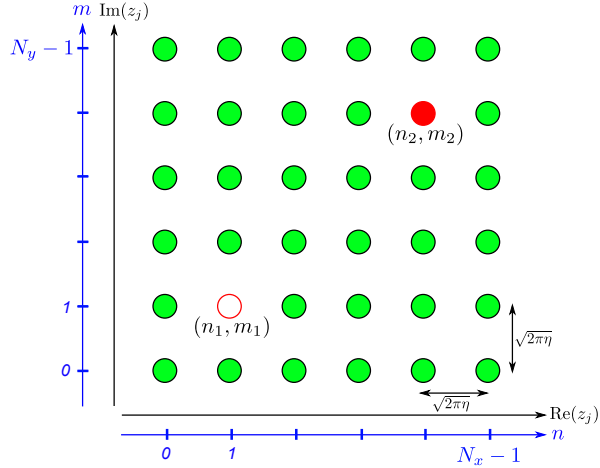}
\caption{Lattice coordinates on the torus. $(n_1,m_1)$ indicates the position of a quasihole, and $(n_2,m_2)$ indicates the position of a quasielectron.}
\label{fig:plotlattice}
\end{indented}
\end{figure}

Let us specifically consider an $N_x \times N_y$ lattice on the torus. We define the coordinates $n \in \{0,1,\ldots,N_x-1\}$ and $m \in \{0,1,\ldots,N_y-1\}$ and take
\begin{equation}
z_{n+mN_x+1}=(n+m i)\sqrt{2\pi\eta},
\end{equation}
such that the area of one lattice site is $2\pi \eta$ (Fig.\ \ref{fig:plotlattice}). In the Landau gauge, the Hamiltonian takes the form
\begin{equation}\label{Ham}
\fl H=-J\sum_{n,m}\left(\hat{d}^\dagger_{n+1,m}\hat{d}_{n,m}e^{-2 \pi i \eta m} + \hat{d}^\dagger_{n,m+1}\hat{d}_{n,m}
+h.c.\right)+U\sum_{n,m}\hat{n}_{n,m}(\hat{n}_{n,m}-1),
\end{equation}
where $\hat{d}$ is the boson annihilation operator, $\hat{n}$ is the number operator, $J$ is the strength of the hopping term, and $U$ is the strength of the interaction term. In the following, we take the limit of hardcore interactions $U\rightarrow \infty$. The amount of magnetic flux going through one plaquette is $\eta$, and in the absence of quasiparticles we fix the number of particles in the system to $\eta N / 2$, so that the number of particles per unit flux is $1/2$.

We want to compare the ground states of \eref{Ham} to the states \eref{statetorus}. For a square lattice, the factor $\prod_{i<j}\thetaf{1/2}{1/2}{\xi_i-\xi_j} {\tau}^{-n_i\eta-n_j\eta}$ is proportional to a Gaussian up to single particle phase factors \cite{torusg}, and in this section we choose $\chi_{n_i}$ such that
\begin{equation}
\prod_i\chi_{n_i}\prod_{i<j}\thetaf{1/2}{1/2}{\xi_i-\xi_j} {\tau}^{-n_i\eta-n_j\eta} \propto
\exp\left(-\frac{1}{2}\sum_i n_i \mathrm{Im}(z_i)^2\right).
\end{equation}
We take $N_x=N_y=6$ and consider the case of $3$ particles on the lattice so that $\eta=1/6$. This ensures magneto-periodic boundary conditions of the Hamiltonian on the torus. For this case $H$ has two ground states, and each of these has an overlap of $0.991$ with a combination of the two Laughlin states on the torus without quasiparticles (we define the overlap of two normalized states $|\psi\rangle$ and $|\psi'\rangle$ as $|\langle\psi|\psi'\rangle|^2$).

We now add a potential to localize a quasihole and a quasielectron excitation by giving an energy penalty to one lattice site to be occupied and an energy penalty to another lattice site to be empty:
\begin{equation}
V=Q \left(\hat n_{n_1,m_1} - \hat n_{n_2,m_2}\right).
\end{equation}
The lattice sites with coordinates $(n_1,m_1)$ and $(n_2,m_2)$ are taken to be as far as possible from each other on the torus (Fig.\ \ref{fig:plotlattice}). We choose here to put the anyons on the lattice sites, since this gives the best overlaps. The number of particles and the number of fluxes are unchanged.

There are two almost degenerate ground states of $H+V$. We compare these two ground states with combinations of the two analytical states on the torus with a quasihole at position $(n_1,m_1)$ and a quasielectron at position $(n_2,m_2)$ and find that they have overlap $0.994$ and $0.989$, respectively, for large $Q$. This shows that the analytical states considered above are relevant for fractional Chern insulator models.

\section{Exact parent Hamiltonians}

It is also possible to derive few-body Hamiltonians for which some of the above states are exact ground states. For $\eta=1$, it can be shown following \cite{contlim} that the operator
\begin{equation}\label{Lambda}
\Lambda_i=\sum_{j(\neq i)}\frac{1}{z_i-z_j} [T_j^{-1}\hat{d}_j-T_i^{-1}\hat{d}_i(q\hat{n}_j-1)]
\end{equation}
annihilates the state $|\psi\rangle_{\vec{0}}$, where $T_k=e^{i\phi_k}e^{-i\pi(k-1)}$ and $\hat{d}_k$ is the hardcore bosonic/fermionic annihilation operator acting on side $k$ for $q$ even/odd. Since $\Lambda_i|\psi\rangle_{\vec{0}}=0$, it follows that $|\psi\rangle_{\vec{0}}$ is a ground state of the Hermitian and positive semi-definite operator $H=\sum_i\Lambda_i^\dag\Lambda_i$. The terms in $H$ are all at most three-body operators. The above derivation does not tell us, whether the ground state of the Hamiltonian $H$ is unique or degenerate. We have therefore computed the ground state of $H$ numerically for several different choices of the lattice obtained by placing the lattice sites at random positions within a region of the plane, and in all cases we find that the ground state is unique if we assume that the number of particles $\sum_in_i$ in the system is fixed. This suggests that the ground state is unique for systems with a fixed number of particles, except possibly for very special choices of the lattice.

Let us now add anyons. We first assume $\sum_ip_i=0$. In this case, the $\delta_n$ factor is unaltered, and we have $|\psi\rangle_{\vec{p}}\propto T'|\psi\rangle_{\vec{0}}$, where $T'=\prod_{i,j}(w_i-z_j)^{p_in_j}$. Therefore $T'\Lambda_iT'^{-1}|\psi\rangle_{\vec{p}}=0$, and it follows that we obtain a Hamiltonian for the state with anyons simply by redefining $T_k=e^{i\phi_k}e^{-i\pi(k-1)}\prod_{i}(w_i-z_k)^{p_i}$. Note that this only changes the strengths of the terms in the Hamiltonian.

In \cite{lp}, it was shown that Hamiltonians can be derived for states with different number of particles by putting a charge at infinity. Utilizing this method for the present case, we can obtain Hamiltonians also for $\sum_i p_i \neq 0$ and $\eta\neq 1$, provided the inequality $\eta-\sum_ip_i/N<1+q/N$ is fulfilled. The Hamiltonian is again given by $H=\sum_i \Lambda_i^\dag \Lambda_i$, but now $T_k=e^{i\phi_k}e^{-i\pi(k-1)}\prod_{i}(w_i-z_k)^{p_i}\prod_j(z_j-z_k)^{1-\eta}$.

The above construction with anyons at particular positions in the ground state is very convenient for manipulating anyons. We can, e.g., start from a state without anyons. By putting a quasielectron and a quasihole coordinate at the same position and pulling them apart, we can create quasielectron-quasihole pairs. Note that this works even when the lattice filling is not $1/2$. We can then do braiding operations, and finally fuse the quasielectrons and quasiholes to vacuum by bringing them to the same point. All these operations can be done by adiabatically changing the interaction strengths in the Hamiltonian.

In studies of lattice models, it is common to require that the anyons are placed on the lattice sites. Here, we allow the anyons to be at any position in the 2D plane. This picture arises naturally from the Hamiltonians derived above. Another possibility for realizing anyons that are not placed on the lattice sites is to introduce (e.g.\ in a cold atoms setting) additional particles of a different type to implement the anyons. These particles are bosons or fermions, but start behaving like anyons when interacting in a suitable way with the system. Experiments along such lines have been proposed in \cite{aux1,aux2,aux3}.

\section{Moore-Read quasielectrons}

The ideas presented above are quite general, since the key point is that the lattice removes the singularity, and are not restricted to Laughlin states. As a further example, we now briefly consider the bosonic Moore-Read state \cite{MR}
\begin{equation}\label{contMR}
\fl |\psi_\textrm{MR}\rangle\propto\int dZ_1 \cdots \int dZ_M \mathrm{Pf}\left(\frac{1}{Z_i-Z_j}\right)
\prod_{i<j}(Z_i-Z_j)\prod_j e^{-|Z_j|^2/4}|Z_1,\ldots,Z_M\rangle,
\end{equation}
where $Z_j$ are the positions of the $M$ particles and $\mathrm{Pf}$ is the Pfaffian. A lattice version of this state can be obtained in a spin-1 system by replacing \eref{vopV} with
\begin{equation}
V_{n_j}=\chi(z_j)^{n_j(2-n_j)}:e^{i(n_j-\eta)\phi(z_j)}:,
\end{equation}
where $\chi$ is the field of a majorana fermion and $n_j\in\{0,1,2\}$ \cite{GTMR,latticeMR}. As demonstrated in \cite{latticeMR}, this state reduces to \eref{contMR} in the continuum limit $\eta\to0^+$ up to some unimportant single particle phase factors, and the topological entanglement entropy of the state remains the same for the whole interval $\eta\in]0,1]$, which suggests that no phase transition occurs when introducing the lattice.

We can obtain wavefunctions with anyons by inserting operators of the form \cite{MR}
\begin{equation}
W_{p_j}=\sigma(w_j) :e^{i\frac{p_j}{2}\phi(w_j)}:,
\end{equation}
where $\sigma$ is the spin operator in the Ising model and $p_j=\pm1$. For the lattice with $\eta=1$, the transformation $\phi\to-\phi$ and $n_j\to2-n_j$ transforms the quasiholes $(p_j=+1)$ into quasielectrons $(p_j=-1)$ and \textit{vice versa}. In particular, $\langle n_j\rangle_{\vec{p}}=2-\langle n_j\rangle_{-\vec{p}}$. The density profiles of quasiholes and quasielectrons are therefore again symmetric around the mean density of one particle per site.

\section{Conclusion}

We have shown that in lattice fractional quantum Hall models it is possible to construct quasielectrons as inverse quasiholes. This leads to quasielectron wavefunctions that are much simpler than previously found quasielectron wavefunctions in the continuum, and we have used this result to make detailed investigations of properties of quasielectrons with Monte Carlo simulations. We have also considered a fractional Chern insulator model and have found that the quasiparticle states constructed describe the fractional Chern insulator states well. Finally, we have found few-body Hamiltonians for which various states containing quasielectrons are exact ground states. These allow creation, braiding, and fusion of anyons by changing the coupling strengths in the Hamiltonian.

The key observation in the above construction is that the inverse of a quasihole does not lead to a singularity in the lattice wavefunctions. This is quite general, and the ideas presented in the present work therefore open up several interesting possibilities for investigating important properties of fractional quantum Hall quasielectrons.

\ack
We would like to thank J. Ignacio Cirac and Germ\'an Sierra for discussions. This work has in part been supported by the EU Integrated Project SIQS and the Villum Foundation.

\section*{References}


\end{document}